\documentclass[superscriptaddress,amsmath,amssymb,11pt,]{revtex4}

\usepackage{amsthm}
\usepackage{charter}
\usepackage{graphicx}
\usepackage{bm}
\usepackage{dsfont}
\usepackage[landscape,papersize={297.1mm,210mm},left=1.9cm,right=1.6cm,top=2.7cm,bottom=2.8cm]{geometry}
\usepackage[pdfstartview=FitH, colorlinks, pdfborder=001, linkcolor=blue, anchorcolor=blue, citecolor=blue, urlcolor=blue]{hyperref}

\begin{document}
\title{Stronger superadditivity relations for multiqubit systems}

\author{Ya-Ya Ren}
\email{yysz7900@163.com}
\affiliation{School of Mathematical Sciences, Capital Normal University, Beijing 100048, China}
\author{Zhi-Xi Wang}
\email{wangzhx@cnu.edu.cn}
\affiliation{School of Mathematical Sciences, Capital Normal University, Beijing 100048, China}
\author{Shao-Ming Fei}
\email{feishm@cnu.edu.cn}
\affiliation{School of Mathematical Sciences, Capital Normal University, Beijing 100048, China}
\affiliation{Max-Planck-Institute for Mathematics in the Sciences, 04103, Leipzig, Germany}

\begin{abstract}
Superadditivity relations characterize the distributions of coherence in multipartite  quantum systems. In this work, we investigate the superadditivity relations related to the $l_1$-norm of coherence $C_{l_1}$ in multiqubit quantum systems. Tighter superadditivity inequalities based on the $\alpha$-th ($\alpha\geqslant 1$) power of $l_1$-norm of coherence are presented for multiqubit states under certain conditions, which include the existing results as special cases. These superadditivity relations give rise to finer characterization of the coherence distributions among the subsystems of a multipartite system. A detailed example is presented.

\end{abstract}

\maketitle

\section{Introduction}
The quantum nature of superposition, entanglement, and measurement are applicable to the quantum information industry. Entanglement is a very unique feature of the quantum sciences and plays a crucial role in quantum information processing. For example, in quantum computing \cite{r1,r7}.
Stemming from the principle of quantum superposition, quantum coherence is another essential feature of quantum mechanics. It also plays an important role in quantum information processing \cite{C1} such as secret sharing \cite{C2}, quantum secure direct communication \cite{r2}, quantum key distribution \cite{r3}, quantum teleportation \cite{r4,r5}, quantum steering \cite{r6}, quantum metrology \cite{C3}, thermodynamics \cite{C4,C5}, and quantum biology \cite{C6}. As a kind of physical resource, recently Baumgaratz $et$ $al.$\cite{C7} proposed a rigorous framework to quantify the coherence. Two intuitive and easily computable measures of coherence are identified, the $l_{1}$-norm of coherence and the relative entropy of coherence. Following this seminal work, various operational measures of quantum coherence have been proposed \cite{C8,C9,C10,C11}. Correspondingly, the dynamics of coherence \cite{C20}, the distillation of coherence \cite{C21,C22} and the relations between quantum coherence and quantum correlations \cite{C23,C24,C25,C26,C27} have been extensively investigated.

The distribution of coherence in multipartite systems is one of the basic problems in the resource theory of coherence. An interesting subject in the theory of coherence is the superadditivity of coherence measures. A given coherence measure $C$ is said to be superadditive if
\begin{equation}
\begin{aligned}
C(\rho_{AB})\geqslant C(\rho_{A})+C(\rho_{B}),\label{1}
\end{aligned}
\end{equation}
for all bipartite density matrices $\rho_{AB}$ of a finite-dimensional system with respect to a particular reference basis $\left \{ |i\rangle_{A}\otimes|j\rangle_{B}\right \}$, where $\rho_{A}=\text{tr}_{B}(\rho_{AB})$ and $\rho_{B}=\text{tr}_{A}(\rho_{AB})$ are the reduced density matrices with respect to the basis $\left\{|i\rangle_{A}\right \}$ and $\left\{|j\rangle_{B}\right \}$, respectively. However, not all coherence measures satisfy such superadditivity relations. The superadditivity for bipartite quantum states based on the relative entropy of coherence has been verified in \cite{C18}. Later, the superadditivity was generalized to the case of tripartite pure states \cite{C16}. A sufficient condition for the convex roof coherence measures to fulfill the superadditivity relations was provided in \cite{C17}. In \cite{C12}, it has been shown that the $l_{1}$-norm of coherence $C_{l_{1}}$ satisfies the superadditivity relations for all multiqubit states. Then the superadditivity of the $l_1$-norm of coherence $C_{l_1}$ for multiqubit systems has been deeply studied \cite{C13,C14}.

In this paper, we show that superadditivity inequalities related to the $\alpha$-th ($\alpha\geqslant 1$) power of $C_{l_1}$ for multiqubit systems can be further improved. A class of tighter superadditivity inequalities in multiqubit systems based on the $\alpha$-th($\alpha \geqslant 1$) power of $l_{1}$-norm of coherence $C_{l_{1}}$ are presented with detailed examples.

\section{Stronger superadditivity relations}
We first recall some basic facts related to the $l_1$-norm of coherence $C_{l_1}$. Let  $\mathcal{H}$ denote a discrete finite-dimensional complex vector space associated with a quantum subsystem. For a quantum state $\rho\in\mathcal{H}$, the $l_{1}$-norm of coherence is given by the sum of the absolute values of the off-diagonal entries of the state $\rho$ \cite{C7},
\begin{equation}
\begin{aligned}
C_{l_{1}}(\rho)=\sum\limits_{i\neq j}|\rho_{ij}|.
\end{aligned}
\end{equation}
The superadditivity relations of the  $l_{1}$-norm of coherence in multiqubit systems has been proved in \cite{C12}.
\begin{equation}
\begin{aligned}
C_{l_{1}}(\rho_{A_{1}A_{2}\cdots A_{n}})\geqslant C_{l_{1}}(\rho_{A_{1}})+C_{l_{1}}(\rho_{A_{2}})+\cdots+C_{l_{1}}(\rho_{A_{n}}).
\end{aligned}
\end{equation}
In \cite{C13}, tighter superadditivity relations in multiqubit systems has been derived,
\begin{equation}
\begin{split}
C_{l_{1}}^{\alpha}(\rho_{A_{1}A_{2}\cdots A_{n}})
&\geqslant C_{l_{1}}^{\alpha}(\rho_{A_{1}})+(2^{\alpha}-1)C_{l_{1}}^{\alpha}
(\rho_{A_{2}})+\cdots+(2^{\alpha}-1)^{m-1}C_{l_{1}}^{\alpha}(\rho_{A_{m}})\\
&\quad +(2^{\alpha}-1)^{m+1}[C_{l_{1}}^{\alpha}(\rho_{A_{m+1}})+\cdots
+C_{l_{1}}^{\alpha}(\rho_{A_{n-1}})]\\
&\quad +(2^{\alpha}-1)^{m}C_{l_{1}}^{\alpha}(\rho_{A_{n}})\label{relation2}
\end{split}
\end{equation}
for all $\alpha\geqslant1$ and $n\geqslant 3$, if for some $m$ ($1\leqslant m\leqslant n-2$) $C_{l_{1}}(\rho_{A_{j}})\leqslant C_{l_{1}}(\rho_{A_{j+1}\cdots A_{n}})$ for $j=m+1, \cdots ,n-1$. Later, the relation (\ref{relation2}) has been improved in \cite{C14},
\begin{equation}\label{5}
	\begin{split}
	C_{l_{1}}^{\alpha}(\rho_{A_{1}A_{2}\cdots A_{n}})
&\geqslant C_{l_{1}}^{\alpha}(\rho_{A_{1}})+\left(\frac{(1+k)^{\alpha}-1}{k^{\alpha}}\right)C_{l_{1}}^{\alpha}(\rho_{A_{2}})+\cdots+\left(\frac{(1+k)^{\alpha}-1}{k^{\alpha}}\right)^{m-1}C_{l_{1}}^{\alpha}(\rho_{A_{m}})\\
&\quad +\left(\frac{(1+k)^{\alpha}-1}{k^{\alpha}}\right)^{m+1}[C_{l_{1}}^{\alpha}(\rho_{A_{m+1}})+\cdots+C_{l_{1}}^{\alpha}(\rho_{A_{n-1}})]\\
&\quad +\left(\frac{(1+k)^{\alpha}-1}{k^{\alpha}}\right)^{m}C_{l_{1}}^{\alpha}(\rho_{A_{n}}),
	\end{split}
\end{equation}
for all $\alpha\geqslant 1$ and $n\geqslant 3$, conditioned that for a real number $k$ ($0<k\leqslant 1$), $C_{l_{1}}(\rho_{A_{i}})\geqslant \frac{1}{k}C_{l_{1}}(\rho_{A_{i+1}\cdots A_{n}})$ for $i=1,2, \cdots, m$, and $C_{l_{1}}(\rho_{A_{j}})\leqslant \frac{1}{k}C_{l_{1}}(\rho_{A_{j+1}\cdots A_{n}})$ for $j=m+1, \cdots, n-1$, $1\leqslant m\leqslant n-2$.

Improving the above results, we have the following theorems.

\textbf{Theorem 1}. Let $k$ and $\delta$ be real numbers satisfying $0<k\leqslant 1$ and $\delta\geq 1$. For any $n$-qubit ($n\geqslant 3$) quantum state $\rho_{A_{1}A_{2}\cdots A_{n}}$ such that, without loss of generality, $C_{l_{1}}(\rho_{A_{i}})\geqslant \frac{1}{k^{\delta}}C_{l_{1}}(\rho_{A_{i+1}\cdots A_{n}})$ for $i=1,2, \cdots, m$, and $C_{l_{1}}(\rho_{A_{j}})\leqslant \frac{1}{k^{\delta}}C_{l_{1}}(\rho_{A_{j+1}\cdots A_{n}})$ for $j=m+1, \cdots, n-1$, $1\leqslant m\leqslant n-2$, we have
	\begin{equation}
	\begin{split}
	C_{l_{1}}^{\alpha}(\rho_{A_{1}A_{2}\cdots A_{n}})
&\geqslant C_{l_{1}}^{\alpha}(\rho_{A_{1}})+\left(\frac{(1+k^{\delta})^{\alpha}-1}{k^{\delta\alpha}}\right)C_{l_{1}}^{\alpha}(\rho_{A_{2}})+\cdots+\left(\frac{(1+k^{\delta})^{\alpha}-1}{k^{\delta\alpha}}\right)^{m-1}C_{l_{1}}^{\alpha}(\rho_{A_{m}})\\
&\quad +\left(\frac{(1+k^{\delta})^{\alpha}-1}{k^{\delta\alpha}}\right)^{m+1}[C_{l_{1}}^{\alpha}(\rho_{A_{m+1}})+\cdots+C_{l_{1}}^{\alpha}(\rho_{A_{n-1}})]\\
&\quad +\left(\frac{(1+k^{\delta})^{\alpha}-1}{k^{\delta\alpha}}\right)^{m}C_{l_{1}}^{\alpha}(\rho_{A_{n}}),\label{theorem10}
	\end{split}
\end{equation}
for all $\alpha\geqslant 1$.

\textbf{Proof}. Due to the superadditivity inequality $C_{l_{1}}(\rho_{AB})\geqslant C_{l_{1}}(\rho_{A})+C_{l_{1}}(\rho_{B})$ for any $2\otimes 2^{n-1}$ bipartite states $\rho_{AB}$ \cite{C13}, and the inequality \cite{C19},
$$
(1+t)^{\alpha}\geqslant 1+\frac{(1+k^{\delta})^{\alpha}-1}{k^{\delta\alpha}}t^{\alpha},
$$
where $k$ and $\delta$ are any real numbers satisfying $0<k\leqslant 1$ and $\delta\geq 1$, $0\leqslant t\leqslant k^{\delta}$ and $\alpha\geqslant 1$, we have
\begin{equation}
\begin{split}
C_{l_{1}}^{\alpha}(\rho_{A_{1}A_{2}\cdots A_{n}})
&\geqslant [C_{l_{1}}(\rho_{A_{1}})+C_{l_{1}}(\rho_{A_{2}\cdots A_{n}})]^{\alpha}\\
&= C_{l_{1}}^{\alpha}(\rho_{A_{1}})\left[1+\frac{C_{l_{1}}(\rho_{A_{2}\cdots A_{n}})}{C_{l_{1}}(\rho_{A_{1}})}\right]^{\alpha}\\
&\geqslant C_{l_1}^{\alpha}(\rho_{A_1})\left\{1+\left(\frac{(1+k^{\delta})^{\alpha}-1}{k^{\delta\alpha}}\right)\left[\frac{C_{l_{1}}(\rho_{A_{2}\cdots A_{n}})}{C_{l_{1}}(\rho_{A_{1}})}\right]^{\alpha}\right\}\\
&= C_{l_1}^{\alpha}(\rho_{A_1})+\left(\frac{(1+k^{\delta})^{\alpha}-1}{k^{\delta\alpha}}\right)C_{l_{1}}^{\alpha}(\rho_{A_{2}\cdots A_{n}})\\
&\geqslant C_{l_1}^{\alpha}(\rho_{A_1})+\left(\frac{(1+k^{\delta})^{\alpha}-1}{k^{\delta\alpha}}\right)C_{l_1}^{\alpha}(\rho_{A_2})+\left(\frac{(1+k^{\delta})^{\alpha}-1}{k^{\delta\alpha}}\right)^{2}C_{l_{1}}^{\alpha}(\rho_{A_{3}\cdots A_{n}})\\
&\geqslant \cdots\\
&\geqslant C_{l_1}^{\alpha}(\rho_{A_1})+\left(\frac{(1+k^{\delta})^{\alpha}-1}{k^{\delta\alpha}}\right)C_{l_1}^{\alpha}(\rho_{A_2})+\cdots+\left(\frac{(1+k^{\delta})^{\alpha}-1}{k^{\delta\alpha}}\right)^{m-1}C_{l_{1}}^{\alpha}(\rho_{A_{m}})\\
&\quad+\left(\frac{(1+k^{\delta})^{\alpha}-1}{k^{\delta\alpha}}\right)^{m}C_{l_{1}}^{\alpha}(\rho_{A_{m+1}\cdots A_{n}}).\label{theorem11}
\end{split}
\end{equation}
Similarly, as $C_{l_{1}}(\rho_{A_{j}})\leqslant \frac{1}{k^{\delta}}C_{l_{1}}(\rho_{A_{j+1}\cdots A_{n}})$ for $j=m+1, \cdots, n-1$, we get
\begin{equation}
\begin{split}
C_{l_{1}}^{\alpha}(\rho_{A_{m+1}\cdots A_{n}})&\geqslant \left(\frac{(1+k^{\delta})^{\alpha}-1}{k^{\delta\alpha}}\right)C_{l_{1}}^{\alpha}(\rho_{A_{m+1}})+C_{l_{1}}^{\alpha}(\rho_{A_{m+2}\cdots A_{n}})\\
&\geqslant \left(\frac{(1+k^{\delta})^{\alpha}-1}{k^{\delta\alpha}}\right)[C_{l_{1}}^{\alpha}(\rho_{A_{m+1}})+\cdots+C_{l_{1}}^{\alpha}(\rho_{A_{n-1}})]+C_{l_{1}}^{\alpha}(\rho_{A_{n}}).\label{theorem12}
\end{split}
\end{equation}
Combining (\ref{theorem11}) and (\ref{theorem12}) we obtain (\ref{theorem10}).
\qed

\textbf{Remark 1}. The Theorem 4 in \cite{C13} is the special case of $k=1$ and $\delta=1$ of our Theorem 1. Our Theorem 1 also includes the Theorem 1 given in \cite{C14} as a special case of $\delta=1$.

In Theorem 1 we have generally assumed that $C_{l_{1}}(\rho_{A_{i}})\geqslant \frac{1}{k^{\delta}}C_{l_{1}}(\rho_{A_{i+1}\cdots A_{n}})$ for $i=1,2, \cdots, m$, and $C_{l_{1}}(\rho_{A_{j}})\leqslant \frac{1}{k^{\delta}}C_{l_{1}}(\rho_{A_{j+1}\cdots A_{n}})$ for $j=m+1, \cdots, n-1$, for some $m$ satisfying $1\leqslant m\leqslant n-2$. In particular, if all $C_{l_{1}}(\rho_{A_{i}})\geqslant \frac{1}{k^{\delta}}C_{l_{1}}(\rho_{A_{i+1}\cdots A_{n}})$ for $i=1,2, \cdots, n-2$, i.e., $m=n-2$, we have the following conclusion:

\textbf{Theorem 2}. If $C_{l_{1}}(\rho_{A_{i}})\geqslant \frac{1}{k^{\delta}}C_{l_{1}}(\rho_{A_{i+1}\cdots A_{n}})$ for all $i=1,2, \cdots, n-2$, then
\begin{equation}\label{t2}
	\begin{split}
	C_{l_{1}}^{\alpha}(\rho_{A_{1}A_{2}\cdots A_{n}})
&\geqslant C_{l_{1}}^{\alpha}(\rho_{A_{1}})+\left(\frac{(1+k^{\delta})^{\alpha}-1}{k^{\delta\alpha}}\right)C_{l_{1}}^{\alpha}(\rho_{A_{2}})+\cdots+\left(\frac{(1+k^{\delta})^{\alpha}-1}{k^{\delta\alpha}}\right)^{n-2}C_{l_{1}}^{\alpha}(\rho_{A_{n-1}})\\
&\quad +\left(\frac{(1+k^{\delta})^{\alpha}-1}{k^{\delta\alpha}}\right)^{m+1}[C_{l_{1}}^{\alpha}(\rho_{A_{m+1}})+\cdots+C_{l_{1}}^{\alpha}(\rho_{A_{n-1}})]\\
&\quad +\left(\frac{(1+k^{\delta})^{\alpha}-1}{k^{\delta\alpha}}\right)^{n-1}C_{l_{1}}^{\alpha}(\rho_{A_{n}}).
	\end{split}
\end{equation}

It is easily verified that our bound (\ref{t2}) is larger than the one from (\ref{5}),
\begin{equation*}
\begin{split}
C^\alpha_{l_1}(\rho_{A_{1}})+\frac{(1+k^\delta)^\alpha-1}{k^{\delta\alpha}}C^\alpha_{l_1}(\rho_{A_{2}})+\Bigg(\frac{(1+k^\delta)^\alpha-1}{k^{\delta\alpha}}\Bigg)^2C^\alpha_{l_1}(\rho_{A_{3}})
&\geq C^\alpha_{l_1}(\rho_{A_{1}})+\frac{(1+k)^\alpha-1}{k^{\alpha}}C^\alpha_{l_1}(\rho_{A_{2}})\\
&+\Bigg(\frac{(1+k)^\alpha-1}{k^\alpha}\Bigg)^2C^\alpha_{l_1}(\rho_{A_{3}})
\end{split}
\end{equation*}
for $\alpha\geq1$.
As an example, let us consider the following three-qubit state,
$$
|\Psi_{A_{1}A_{2}A_{3}}\rangle=\frac{|0\rangle+|1\rangle}{\sqrt{2}}\otimes |0\rangle\otimes \frac{|0\rangle+3|1\rangle}{\sqrt{10}}.
$$
For $|\Psi_{A_{1}A_{2}A_{3}}\rangle$ we have $C_{l_1}(\rho_{A_{1}})=1$, $C_{l_1}(\rho_{A_{2}})=0$, $C_{l_1}(\rho_{A_{3}})=\frac{3}{5}$, $C_{l_1}(\rho_{A_{2}A_{3}})=\frac{3}{5}$. Hence, we can choose $\delta=2$ and $k=\frac{4}{5}$. We have the lower bound of (\ref{t2}),
\begin{equation*}
y_1\equiv C^\alpha_{l_1}(\rho_{A_{1}})+\frac{(1+k^\delta)^\alpha-1}{k^{\delta\alpha}}C^\alpha_{l_1}(\rho_{A_{2}})+\Bigg(\frac{(1+k^\delta)^\alpha-1}{k^{\delta\alpha}}\Bigg)^2C^\alpha_{l_1}(\rho_{A_{3}})
=1+[(\frac{41}{25})^\alpha-1]^2(\frac{375}{256})^\alpha.
\end{equation*}
While the lower bound of (\ref{5}) is given by
\begin{equation*}
y_2\equiv C^\alpha_{l_1}(\rho_{A_{1}})+\frac{(1+k)^\alpha-1}{k^{\alpha}}C^\alpha_{l_1}(\rho_{A_{2}})+\Bigg(\frac{(1+k)^\alpha-1}{k^\alpha}\Bigg)^2C^\alpha_{l_1}(\rho_{A_{3}})
=1+[(\frac{9}{5})^\alpha-1]^2(\frac{15}{16})^\alpha.
\end{equation*}
Fig. 1 shows that our result is indeed tighter than the one given in \cite{C14}.

\begin{figure}
  \centering
  \includegraphics[width=9cm]{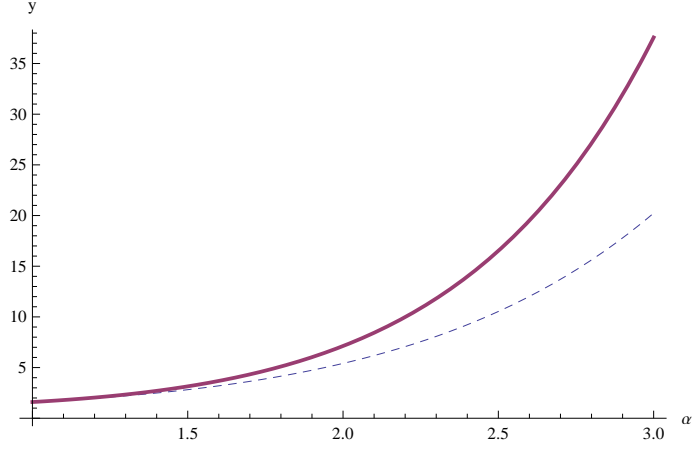}\\
  \caption{ The $l_1$-norm of coherence $C_{l_1}$ with respect to $\alpha$:
  the solid line is for $y_1$ and the dashed line for $y_2$ from the result in \cite{C14}.}
\end{figure}

Inequality (\ref{theorem10}) can be further generalized to the following theorem, with a
similar proof to (\ref{theorem10}).

\textbf{Theorem 3}. Let $k$, $\delta$ and $\beta$ be real numbers with $0<k\leqslant 1$ and  $\delta,\beta\geqslant 1$. For any $n$-qubit quantum state such that $C_{l_{1}}^{\beta}(\rho_{A_{i}})\geqslant \frac{1}{k^{\delta}}C_{l_{1}}^{\beta}(\rho_{A_{i+1}\cdots A_{n}})$ for $i=1,2, \cdots, m$, and $C_{l_{1}}^{\beta}(\rho_{A_{j}})\leqslant \frac{1}{k^{\delta}}C_{l_{1}}^{\beta}(\rho_{A_{j+1}\cdots A_{n}})$ for $j=m+1, \cdots, n-1$, $1\leqslant m\leqslant n-2$ and $n\geqslant 3$, we have
\begin{equation}
\begin{split}
C_{l_{1}}^{\alpha\beta}(\rho_{A_{1}A_{2}\cdots A_{n}})
&\geqslant C_{l_{1}}^{\alpha\beta}(\rho_{A_{1}})
+\left(\frac{(1+k^{\delta})^{\alpha}-1}{k^{\delta\alpha}}\right)
C_{l_{1}}^{\alpha\beta}(\rho_{A_{2}})+\cdots
+\left(\frac{(1+k^{\delta})^{\alpha}-1}{k^{\delta\alpha}}\right)^{m-1}
C_{l_{1}}^{\alpha\beta}(\rho_{A_{m}})\\
&\quad +\left(\frac{(1+k^{\delta})^{\alpha}-1}{k^{\delta\alpha}}\right)^{m+1}
[C_{l_{1}}^{\alpha\beta}(\rho_{A_{m+1}})+\cdots+C_{l_{1}}^{\alpha\beta}(\rho_{A_{n-1}})]\\
&\quad +\left(\frac{(1+k^{\delta})^{\alpha}-1}{k^{\delta\alpha}}\right)^{m}
C_{l_{1}}^{\alpha\beta}(\rho_{A_{n}})
\label{theorem31}
\end{split}
\end{equation}
for all $\alpha\geqslant 1$.

\textbf{Remark 2}. Theorem 3 reduces to Theorem 1 when $\beta=1$. In particular, when $m=n-2$
Theorem 3 gives rise to a simpler stronger superadditivity relation:

\textbf{Theorem 4}. If $C_{l_{1}}^{\beta}(\rho_{A_{i}})\geqslant \frac{1}{k^{\delta}}C_{l_{1}}^{\beta}(\rho_{A_{i+1}\cdots A_{n}})$ for all $i=1,2, \cdots, n-2$, we have
	\begin{equation}
	\begin{split}
	C_{l_{1}}^{\alpha\beta}(\rho_{A_{1}A_{2}\cdots A_{n}})
&\geqslant C_{l_{1}}^{\alpha\beta}(\rho_{A_{1}})
+\left(\frac{(1+k^{\delta})^{\alpha}-1}{k^{\delta\alpha}}\right)
C_{l_{1}}^{\alpha\beta}(\rho_{A_{2}})+\cdots
+\left(\frac{(1+k^{\delta})^{\alpha}-1}{k^{\delta\alpha}}\right)^{n-2}
C_{l_{1}}^{\alpha\beta}(\rho_{A_{n-1}})\\
&\quad +\left(\frac{(1+k^{\delta})^{\alpha}-1}{k^{\delta\alpha}}\right)^{n-1}
C_{l_{1}}^{\alpha\beta}(\rho_{A_{n}})\label{theorem41}
	\end{split}
\end{equation}
for all $\alpha\geqslant 1$.

Note that not all coherence measures satisfy a superadditivity relation like the inequality (\ref{1}) for all quantum states. The method used in Theorem 4 can be applied to derive tighter superadditivity inequalities for the $\alpha$-th ($\alpha\geqslant 1$) power of coherence measures satisfying the superadditivity relation.

\section{Conclusion}
Superadditivity relation is a fundamental property with respect to multipartite quantum systems. In this paper, we have focused on the distributions of quantum coherence characterized by the superadditivity relations. We have proposed a class of tighter superadditivity inequalities related to the $\alpha$-th ($\alpha\geqslant 1$) power of the $l_1$-norm coherence $C_{l_1}$ for multiqubit systems. These new inequalities give rise to finer characterizations for the coherence distribution. Our results provide better understanding of multipartite coherence and may highlight related researches on other quantum coherence measures.

\begin{acknowledgments}
This work is supported by NSF of China under No. 12075159, Key Project of Beijing Municipal Commission of Education (KZ201810028042), Beijing Natural Science Foundation (Z190005), Academy for Multidisciplinary Studies, Capital Normal University, and Shenzhen Institute for Quantum Science and Engineering, Southern University of Science and Technology, Shenzhen 518055, China (No. SIQSE202001).
\end{acknowledgments}

\end{document}